\documentclass[aps,prl,twocolumn,superscriptaddress,floatfix,nofootinbib,longbibliography]{revtex4-2}
\usepackage{amsmath,amssymb,bm,braket}
\usepackage{graphicx}
\usepackage{xcolor}
\usepackage[colorlinks=true,allcolors=blue]{hyperref}

\newcommand{\kk}{\mathbf{k}}
\newcommand{\GG}{\mathbf{G}}
\newcommand{\KK}{\mathbf{K}}
\newcommand{\rr}{\mathbf{r}}

\newcommand{\AAop}{\hat{\mathbf{A}}}

\newcommand{\ah}{\hat a}
\newcommand{\adag}{\hat a^{\dagger}}
\newcommand{\eps}{\bm{\varepsilon}}

\newcommand{\Vcav}{V_{\mathrm{cav}}}

\begin{document}

\title{Plane-Wave Photon-Fock Cavity QED-DFT: Chiral-Cavity-Induced Topology
in Graphene}

\author{Yetmgeta Aklilu}
\affiliation{Department of Physics and Astronomy, Vanderbilt
University, Nashville, Tennessee, 37235, USA}
\author{K\'alm\'an Varga}
\email{kalman.varga@vanderbilt.edu}
\affiliation{Department of Physics and Astronomy, Vanderbilt
University, Nashville, Tennessee, 37235, USA}

\begin{abstract}
We develop an explicit plane-wave (PW) $\times$ photon-Fock
approach to the quantized, velocity-gauge Pauli--Fierz Hamiltonian
for periodic, first-principles calculations: because the quantized
vector potential is spatially uniform, the light--matter coupling
reduces to an operator-valued shift of the crystal momentum,
$\KK\to\KK+\hat{\mathbf A}/c$, and the existing plane-wave machinery
of ordinary solid-state DFT is reused essentially unchanged. This
PW$\times$Fock construction puts first-principles cavity QED of
periodic solids on the same footing as the Fock-space
coupled-cluster and configuration-interaction methods already
developed for molecules, opening hitherto inaccessible
systems---materials in a linear or chiral cavity, in particular---to
first-principles plane-wave calculations. We illustrate the method
with monolayer graphene in linear and chiral cavities, obtaining a
polarization-selective Haldane gap together with the corresponding
density-of-states, real-space, and circular-dichroism optical
signatures. A full Brillouin-zone Berry-curvature calculation confirms
the topological character of the chiral-cavity gap directly: the
occupied manifold carries a quantized Chern number that steps through
a non-monotonic sequence, $C=1\to3\to{-1}\to1\to2\to{-1}\to1$, with
two narrow, sign-reversed windows---a property of extended,
vacuum-dressed matter with no counterpart in a finite molecular
system.
\end{abstract}

\maketitle

Coupling matter to the vacuum field of an optical cavity, without any
external drive, has emerged as one of the most active frontiers of
condensed-matter and chemical physics: experiments report
cavity-induced changes to transport, chemical reactivity,
superconductivity, intermolecular energy transfer, and the electronic
structure of two-dimensional itinerant
systems~\cite{Ebbesen2016,Ruggenthaler2018,GarciaVidal2021,
schlawin2022,paravicini2019,appugliese2022,thomas2016,Thomas2019,Shalabney2015,
Nagarajanandothers2021,Galego2015,Galegoandothers2018,Schafer2019,Weidman2024,
Liebenthal2024,Fregoniandothers2018,Galego2020,KnaCohenandothers2019,
Ribeiroandothers2018,HerreraandSpano2019,Feistandothers2017,Haugland2021,
Feist2015,Coles2014,Schachenmayer2015,Mandal2020b,2fw2-lbhy},
and a large theoretical literature has developed around such
vacuum-dressed phases~\cite{Basov2025,Mazza2019,Ruggenthaler2023}. A
recurring theme is that the cavity vacuum can act as a
symmetry-breaking or topology-inducing agent: chiral (circularly
polarized) cavities have been proposed to break time-reversal symmetry
and generate Haldane-type gaps and nontrivial band topology in
two-dimensional Dirac materials~\cite{wang2019,hubener2021,dag2024},
recently confirmed at the first-principles QEDFT level in the
three-dimensional topological material HgTe~\cite{Shin2026sciadv},
with related proposals for the quantum Hall
effect~\cite{1tlw-g26r} and chiral spin liquids in frustrated lattice
magnets~\cite{8qx2-xxh2}---an effect intrinsic to the extended,
periodic band structure rather than to a finite cluster or molecule.

The theoretical description of these effects requires a
first-principles framework that treats electrons and photons on the
same footing. Quantum electrodynamical density-functional theory
(QEDFT) provides such a
framework~\cite{Ruggenthaler2014,Flick2015,Flick2018,Tokatly2013,Tokatly2018,Novokreschenov2023,Pellegrini2015},
complemented by purpose-built exchange-correlation
functionals~\cite{Lu2024,Tasci2025,Ahmadabadi2025,MejiaRodriguez2025},
reduced-density-matrix and photon-free
reformulations~\cite{Buchholz2019,Schafer2021}, and, for finite
systems, wavefunction-based routes---coupled-cluster, and
configuration-interaction
formulations~\cite{Flick2017,Flick2020,Malave2022,Aklilu2026,Haugland2020,Riso2022,Mordovina2020,
Vu2024,Vu2025,Matousek2024,Pavosevic2021,Warren2024,Hassan2024,Sheng2025,Chiari2025,Bauman2026}.
In practice, QEDFT is most often realized in a real-space or
localized-basis representation suited to molecules and clusters

For extended, periodic systems, however, the natural and by far most
developed electronic-structure technology is the plane-wave (PW)
pseudopotential method~\cite{Bloch1928,Herring1940,IhmZungerCohen1979}
built on density functional
theory~\cite{Hohenberg1964,Kohn1965}. PW codes underpin the majority
of production DFT calculations of solids~\cite{Payne1992,Martin2004},
implemented in packages such as VASP, Quantum ESPRESSO, ABINIT,
CASTEP, Qbox, PWmat, and
JDFTx~\cite{Kresse1993,Kresse1996a,Kresse1996b,Giannozzi2009,Giannozzi2017,
Gonze2009,Clark2005,Gygi2008,Jia2013,Jia2016,Sundararaman2017}:
they are variationally systematic in a single convergence parameter,
exploit the FFT for $O(N\log N)$ scaling, and interface naturally with
pseudopotential/projector-augmented-wave
machinery~\cite{Vanderbilt1990,Blochl1994,Troullier1991,Kresse1999}.

We extend these plane-wave calculations to QED cavities: using a
product basis of plane waves for the matter and a Fock basis for the
light, the Pauli--Fierz Hamiltonian in the long-wavelength limit,
combined with the density-functional-theory Hamiltonian, can be
solved with nearly the same simplicity as conventional plane-wave DFT,
simply by replacing $\KK\to\KK+\AAop/c$, together with an extension of
the nonlocal-pseudopotential gauge correction to this quantized field
(Supplemental Material~\cite{supp}). This
plane-wave$\,\times\,$photon-Fock product ansatz has already been used
successfully for molecules~\cite{Malave2022,Aklilu2026,42x4-nlwy}, and
it is the same electron-photon Fock-space construction underlying the
wavefunction-based QED many-body methods developed for molecular
systems: QED coupled-cluster theory, notably by Koch and
coworkers~\cite{Haugland2020,Lexander2024}, by DePrince and
coworkers~\cite{DePrince2021,Liebenthal2023}, and by
others~\cite{Mordovina2020,Bauman2026}, and QED
configuration-interaction theory, including work by Govind and
coworkers~\cite{Vu2024,Vu2025}. The present work extends this same
product-basis construction, for the first time, to a periodic,
first-principles plane-wave pseudopotential setting, putting
first-principles cavity QED of extended solids on the same conceptual
footing---and, in principle, the same systematic path to higher-order
electron-photon correlation---as these established Fock-space
many-body approaches for molecules. Because the cavity photon is
thereby carried exactly, at the level of each polaritonic orbital, in
the same momentum-space language as the electrons rather than added
as a perturbative or classical correction, this construction opens
periodic, first-principles cavity-QED calculations to properties of
vacuum-dressed matter that were inaccessible before---most concretely,
a genuine Brillouin-zone topological invariant, which we show below
undergoes a cavity-induced, quantized Chern-number staircase in a
two-dimensional Dirac material. We illustrate the method with a
first-principles study of monolayer graphene in linear and chiral
cavities, and show the two are symmetry-complementary: the chiral
cavity breaks time-reversal symmetry, opening a Haldane gap with a
quantized Chern number and circular dichroism, while the linear cavity
preserves time reversal but breaks the lattice's threefold rotational
symmetry, leaving the Dirac point gapless yet optically
anisotropic---together with their density-of-states and real-space
signatures.

Using a plane-wave DFT code to host a QEDFT calculation is, by itself,
not new: electron--photon coupling has recently been implemented
within a plane-wave pseudopotential code and extended to phonons and
real-time dynamics~\cite{Fan2026unified}. That treatment, however,
couples the photon through an effective, density-functional-level
electron--photon interaction rather than through an explicit
photon-Fock product basis: it does not, to our knowledge, carry
polaritonic orbitals or a Fock-space ansatz of the kind used in QED
coupled-cluster theory, QED configuration-interaction theory, or the
present work.

\textit{Method.}---We work in Hartree atomic units. The
nonrelativistic Pauli--Fierz Hamiltonian in Coulomb gauge, restricted
to a single cavity mode of frequency $\omega$ in the long-wavelength
(dipole) approximation~\cite{Fierz1939,Rokaj2018}, reads
\begin{equation}
\hat{H} = \sum_{i}\frac{1}{2}\left(-i\nabla_{i}+\frac{1}{c}\AAop\right)^{2}
  + \hat{V}_{\mathrm{ext}} + \hat{W}_{ee}
  + \omega\!\left(\adag\ah+\tfrac{1}{2}\right),
\label{eq:PF}
\end{equation}
where the sum runs over electrons $i$, $\hat V_{\mathrm{ext}}$ and
$\hat W_{ee}$ are the usual external and electron--electron potentials,
and the last term is the bare photon energy. The quantized,
spatially uniform vector potential is
\begin{equation}
\AAop = A_{0}\left(\eps\,\ah+\eps^{*}\adag\right),
\qquad
A_{0}=c\sqrt{\frac{2\pi}{\Vcav\,\omega}} ,
\label{eq:Aop}
\end{equation}
with $\ah,\adag$ the photon annihilation and creation operators,
$\Vcav$ the mode (quantization) volume, and the polarization vector
$\eps$ kept complex from the outset so that circular (chiral) modes
are included without modification. Because $\AAop$ carries no spatial
dependence, it commutes with all lattice translations, so the crystal
momentum $\kk$ remains a good quantum number---the structural fact the
whole method rests on. (The alternative length-gauge Hamiltonian
contains a dipole self-energy $(\bm\lambda\cdot\rr)^2$ that is
ill-defined under periodic boundary conditions, which makes the
velocity gauge of Eq.~\eqref{eq:PF} particularly natural for a direct
plane-wave treatment of periodic systems, avoiding the subtleties
associated with defining the dipole operator under periodic boundary
conditions.)

For each $\kk$ we expand the polaritonic Kohn--Sham orbitals in the
tensor-product basis of plane waves and photon-number (Fock) states,
\begin{equation}
\Psi_{n\kk}(\rr) = \sum_{\GG}\sum_{m=0}^{M}
  C_{n\kk}(\GG,m)\,\frac{1}{\sqrt{\Omega}}\,e^{i(\kk+\GG)\cdot\rr}\,\ket{m},
\label{eq:ansatz}
\end{equation}
with reciprocal-lattice vectors $\GG$, cell volume $\Omega$, and a
Fock-space cutoff $M$; the coefficients $C_{n\kk}(\GG,m)$ describe
a nonseparable light--matter structure within each polaritonic orbital
and are not, in general, a single factorized product. Acting on this basis with
$\KK\equiv\kk+\GG$, the kinetic operator in Eq.~\eqref{eq:PF} becomes
diagonal in $\GG$ and simply replaces $\KK\to\KK+\AAop/c$ everywhere it
appears---the entire content of the light--matter coupling. In the
basis $(\GG,m)$ the Kohn--Sham Hamiltonian at fixed $\kk$ takes the
block form
\begin{equation}
H^{(\kk)}_{\GG m,\GG'm'}
= v_{\mathrm{KS}}^{\mathrm{loc}}(\GG-\GG')\,\delta_{mm'}
+ \delta_{\GG\GG'}\,h^{(\KK)}_{mm'}
+ \big[V^{\mathrm{NL}}\big]^{(\kk)}_{\GG m,\GG'm'},
\label{eq:Hblock}
\end{equation}
where $v_{\mathrm{KS}}^{\mathrm{loc}}$ is the ordinary local
Kohn--Sham potential (diagonal in the photon index $m$), $h^{(\KK)}_{mm'}$
is a small, $\GG$-diagonal photonic block, and $V^{\mathrm{NL}}$ is the
nonlocal pseudopotential discussed below. Unlike the local term,
$V^{\mathrm{NL}}$ is not diagonal in $m$: its kernel is finite-range
rather than local, so the minimal-coupling gauge phase acts on
$\rr\neq\rr'$ within that range, where it is a genuine photon operator
rather than the identity (Sec.~S2 of the Supplemental
Material~\cite{supp}). For linear polarization
($\eps$ real) it is
\begin{align}
h^{(\KK)}_{mm'} =\;&
\Big[\tfrac{1}{2}|\KK|^{2}+\omega(m+\tfrac12)
   +\tfrac{A_0^2}{2c^2}(2m{+}1)\Big]\delta_{mm'}
\nonumber\\
&+\frac{A_0}{c}(\KK\!\cdot\!\eps)
   \Big[\sqrt{m}\,\delta_{m',m-1}+\sqrt{m{+}1}\,\delta_{m',m+1}\Big]
\nonumber\\
&+\frac{A_0^2}{2c^2}
   \Big[\sqrt{m(m{-}1)}\,\delta_{m',m-2}
\nonumber\\
&\qquad\quad+\sqrt{(m{+}1)(m{+}2)}\,\delta_{m',m+2}\Big].
\label{eq:hphot}
\end{align}
The first line is the shifted kinetic energy plus bare photon energy;
the second is the paramagnetic (Jaynes--Cummings-like) coupling~\cite{Jaynes1963,TavisCummings1968}, 
linear in $\AAop$, which connects adjacent photon sectors
$\Delta m=\pm1$; the third is the diamagnetic term, quadratic in
$\AAop$, which connects $\Delta m=0,\pm2$. This complementary
sparsity---local potential diagonal in Fock space, light--matter
coupling diagonal in $\GG$---means applying $\hat H$ costs only
$M{+}1$ FFTs for the local potential plus a banded operation for the
photonic part; the only new physical input is $A_0\eps$ itself, and
the existing FFT and pseudopotential infrastructure of a PW code is
reused essentially unchanged.

The polarization vector $\eps$ enters Eq.~\eqref{eq:hphot} directly and
controls a qualitative difference between the linear and chiral drives
studied below: for circular polarization $\eps\cdot\eps=0$ identically
and the diamagnetic ($\Delta m=\pm2$) term vanishes, while for linear
polarization it survives and renormalizes the effective cavity
frequency. Both drives nonetheless couple every plane-wave component
identically, so this polarization dependence costs nothing extra in
the implementation---it is carried automatically by Eq.~\eqref{eq:hphot}.

The minimal-coupling substitution correctly gauges the kinetic and
local terms but not the nonlocal (Kleinman--Bylander) pseudopotential
$V^{\mathrm{NL}}$ in Eq.~\eqref{eq:Hblock}; we shift it consistently,
$\beta(\KK)\to\beta(\KK+\AAop/c)$, following the classical
velocity-gauge correction of Ismail-Beigi, Chang, and
Louie~\cite{ismail2001}, extended here to the quantized field. The
full gauge-consistent treatment is given in the Supplemental
Material~\cite{supp}.

Separately, when the coupled problem is solved at the mean-field
(Kohn--Sham) level with a factorized ansatz per orbital, the Pauli
exclusion principle in the matter--photon channels can be imposed via
a projector $\hat Q$ that removes two artifacts of the factorized
ansatz: an unphysical self-displacement channel, in which a
current-carrying Bloch state couples to its own photon replica and
spuriously inflates the photon number, and spurious resonances between
pairs of occupied bands mediated by a virtual photon, a process
genuinely Pauli-blocked in the many-body problem; the full construction
of $\hat Q$ is given in the Supplemental Material~\cite{supp}.
Although this projector removes the relevant channels at weak
coupling, rebuilding it self-consistently from the instantaneous
occupied manifold produces an occupation--projection feedback
instability at stronger coupling. The production calculations reported
below therefore use the stably converged, unprojected treatment; the
Supplemental Material~\cite{supp} quantifies the resulting limitation
and shows that the Haldane gap changes only by a few percent when the
projector can be applied.

Two limitations of the present implementation are worth stating before
turning to the numerical example. First, only a single cavity mode is
retained; additional modes can be added in principle, but the
Fock-space dimension grows combinatorially with the number of modes,
so a naive extension is costly. An efficient truncation scheme that
tames this growth has been introduced and tested in
Ref.~\cite{PhysRevA.110.043119}, and will allow the present method to
be extended to several modes without the full combinatorial cost.
Second, the electron--photon system is treated with a matter-only
exchange-correlation functional throughout: the single-mode photon
itself is treated exactly at the level of each polaritonic orbital,
but explicit photon exchange-correlation---correlation between the
electronic and photonic degrees of freedom beyond this mean-field
treatment---is neglected. A simple, approximate scheme for including
it is discussed in the Supplemental Material~\cite{supp} and will be
used in future work, but it is not included in any of the numerical
results presented below; the Supplemental Material~\cite{supp} also
discusses a useful benchmark for what such a term would need to
capture.

\textit{Graphene in a cavity.}---Our numerical example is monolayer
graphene coupled to a single cavity mode of either linear or chiral
(circular) polarization: the prototypical two-dimensional Dirac
material and the system in which chiral-cavity band-gap engineering
has been most
studied~\cite{wang2019,hubener2021,Yang2025,Tay2025,dag2024,M.Kulkarni2026}
including with a recent
photon-free self-consistent Hartree--Fock treatment of the same
polarization dependence~\cite{Liu2025sciadv},
and recent experimental progress toward realizing chiral cavities with
genuinely broken time-reversal symmetry in the
terahertz~\cite{M.Kulkarni2026,Tay2025}. In the calculations
we use a rectangular four-atom graphene cell with cavity frequency
$\omega=2$~eV and coupling strengths $A_0/c=0.01$--$0.05$~a.u.---numerically
modest in these units, but corresponding, once translated into a vacuum
field and mode volume, to a regime extending from near the conventional
ultrastrong-coupling threshold ($g/\omega\approx0.07$) to
$g/\omega\approx0.33$, approaching but not reaching the deep-strong
regime. The implied cavity mode volumes are correspondingly small,
well beyond what present-day dielectric or plasmonic cavities achieve,
so these results are best read as an idealized theoretical benchmark
rather than a near-term experimental target. Further computational
details, including this translation, are given in the Supplemental
Material~\cite{supp}.

\begin{figure*}[t]
\centering
\includegraphics[width=0.85\textwidth]{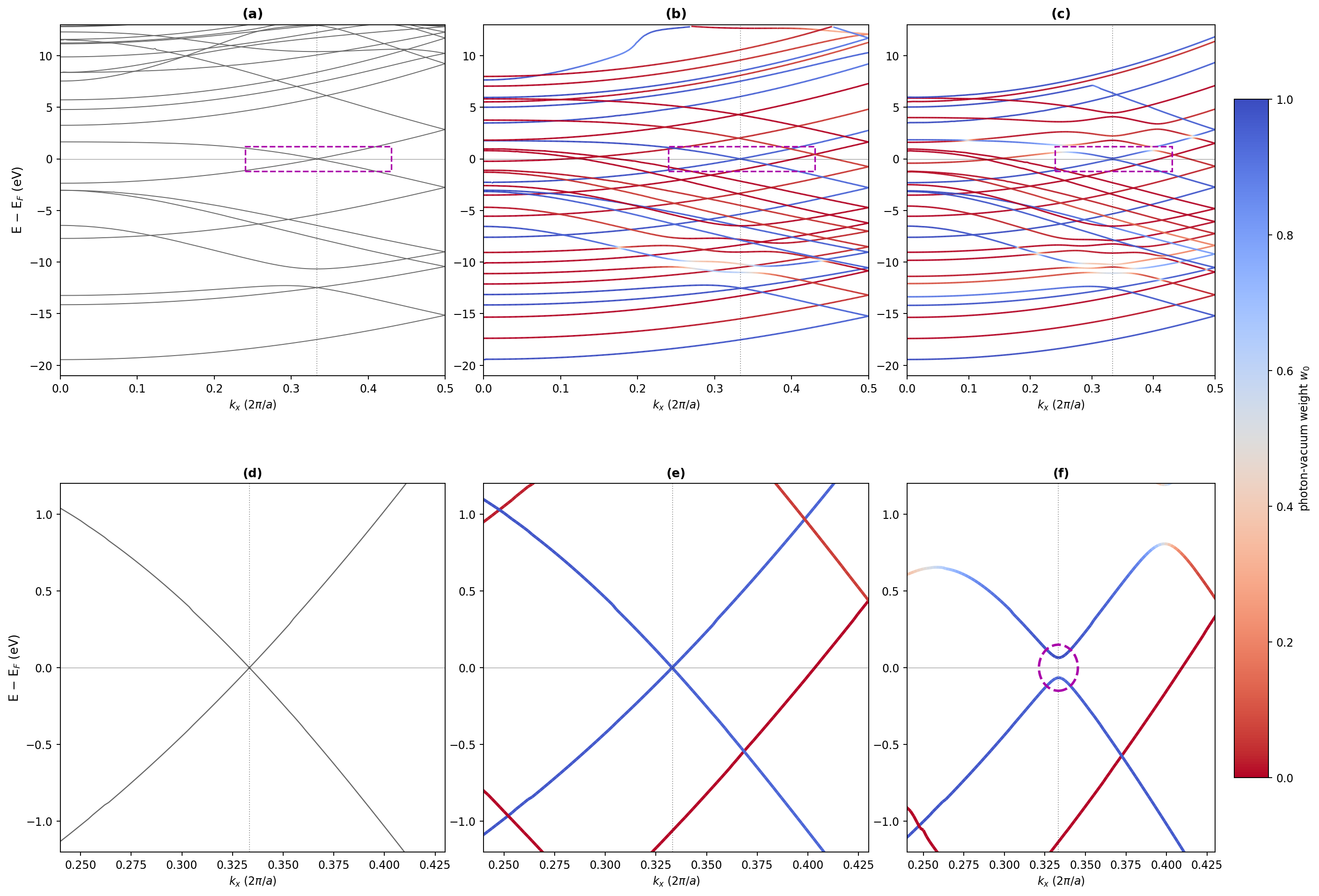}
\caption{Polariton band structure at intermediate coupling
($A_0/c=0.04$, $\omega=2$~eV, $M=3$), each panel referenced to its own
Fermi level. (a)--(c): full-window bands, cavity-free/linear/chiral.
(d)--(f): Dirac-point zoom. Color is $w_0$ [Eq.~\eqref{eq:w0}]; blue
$1$, matter-like; red $0$, photon-dressed. The linear cavity leaves
the Dirac crossing gapless (e), as symmetry requires (see text); the
chiral (Haldane) gap ($129$~meV, dashed ellipse, f) opens with the branches still
matter-like at $\KK$, the real hybridization occurring off-$\KK$ where
a different photon replica crosses the cone.}
\label{fig:bands}
\end{figure*}

Figure~\ref{fig:bands} shows the band structure along the $\Gamma$--$X$
direction at $A_0/c=0.04$, together with a Dirac-point zoom, for the
cavity-free, linear-cavity, and chiral-cavity cases. In the
rectangular cell the two inequivalent Dirac points of graphene fold
onto the $\Gamma$--$X$ line at $k_x=\frac13$ ($2\pi/a$); the folded
Dirac crossing is clearly visible in the cavity-free panel. Each band
spawns photon replicas displaced by multiples of $\omega$, colored by
their photon-vacuum weight
\begin{equation}
w_{0,n\kk}=\frac{\sum_\GG|C_{n\kk}(\GG,0)|^2}{\sum_{\GG,m}|C_{n\kk}(\GG,m)|^2}\in[0,1],
\label{eq:w0}
\end{equation}
the fraction of the polariton norm residing in the zero-photon sector;
$w_0\to1$ marks an essentially undressed matter state. Bands are
connected across $\kk$ by maximum wavefunction overlap between
neighboring $\kk$-points rather than re-sorted by energy at each
$\kk$, so that a single color trace follows one physical state
continuously through a crossing instead of jumping between branches
whenever two polariton replicas happen to become quasidegenerate.

\begin{table}[t]
\centering
\caption{Chiral (Haldane) gap vs.\ coupling, $18\times12$ mesh, and the
implied Jaynes--Cummings coupling $\lambda=\sqrt{\Delta(\Delta+\omega)}$.}
\begin{tabular}{ccc}
\hline\hline
$A_0/c$ & $\Delta_{\mathrm{Hal}}$ (meV) & $\lambda/A_0$ (meV) \\
\hline
0.01 & 9   & 13447 \\
0.02 & 34  & 13149 \\
0.03 & 75  & 13150 \\
0.04 & 129 & 13102 \\
0.05 & 193 & 13012 \\
\hline\hline
\end{tabular}
\label{tab:scaling}
\end{table}

Figure 1 shows that the chiral cavity opens up the Dirac-point gap
while the linear cavity does not. At $A_0/c=0.04$ the linear cavity
leaves the Dirac crossing gapless to numerical precision (e), with
$w_0$ above $0.93$ for both branches everywhere in the zoomed window:
two essentially undressed matter states meeting at an unperturbed
crossing. This null result is required by symmetry, not merely
observed: $\AAop$ is spatially uniform and couples identically to the
two sublattices, so it cannot generate a trivial (Semenoff-type)
mass~\cite{Semenoff1984},
while the topological (Haldane-type) mass generated at second order in
the light--matter coupling is odd in the polarization's handedness,
$\chi\equiv i(\eps\times\eps^*)\!\cdot\!\hat{\mathbf
z}=2\,\mathrm{Im}(\varepsilon_x^*\varepsilon_y)$: $\chi$ vanishes
identically for any real (linearly polarized) $\eps$, since
$\eps^*=\eps$ makes $\eps\times\eps^*=0$, forbidding the mass for the
linear cavity as well, and it reverses sign under $\eps\to\eps^*$---the
polarization's response to time reversal---consistent with a Haldane
mass being itself odd under time reversal (Supplemental
Material~\cite{supp} gives the full symmetry argument). The chiral
cavity, where $\chi=\pm1\neq0$ for the ideal circular polarizations
used here and time-reversal symmetry is genuinely broken, opens
$\Delta_{\mathrm{Hal}}=129$~meV at the same coupling, the same
polarization dependence traced across the full coupling range in
Table~\ref{tab:scaling}. The character of that gap is worth making
explicit: at the Dirac point both branches remain predominantly
matter-like ($w_0\gtrsim0.93$), so the mass gap itself does not
require strong photon dressing to open---it is generated in the
matter sector by virtual, off-resonant coupling to the full band
manifold. This is not a special feature of one coupling: repeating
the analysis at every point in Table~\ref{tab:scaling}, $w_0$ at $\KK$
for the lower branch decreases smoothly and monotonically from
$0.995$ at $A_0/c=0.01$ to $0.901$ at $A_0/c=0.05$, while the upper
branch stays above $0.995$ throughout, so the matter character of the
gap-forming states at $\KK$ is a property of the whole coupling range
studied, not an artifact of a particular point.

Table~\ref{tab:scaling} gives the chiral (Haldane) gap on this mesh at
each coupling studied. The Jaynes--Cummings (JC) form for a saturating
two-level doublet, $\Delta=\sqrt{(\omega/2)^2+\lambda^2}-\omega/2$,
fits all five points to within $3\%$ in the implied coupling
$\lambda/A_0=\sqrt{\Delta(\Delta+\omega)}/A_0$: a single,
coupling-independent proportionality $\lambda\propto A_0$ feeds the
saturating doublet consistently across the entire range studied. Because
$\Delta/\omega\lesssim0.1$ over this range, a natural concern is that
this good fit does not yet distinguish genuine JC saturation from its
leading quadratic limit, $\Delta\simeq\lambda^2/\omega$; a direct
least-squares comparison, however, favors the full JC form appreciably
(worst-point residual $5.7\%$ versus $12.5\%$ for the best-fit quadratic
alone; Supplemental Material~\cite{supp}), evidence that the effective
light--matter coupling underlying the Haldane mass is already better
described by the saturating two-level picture than by leading-order
perturbation theory across the whole range studied, though extending
the coupling further would make the distinction unambiguous.

Genuine light--matter hybridization is present at every coupling too,
but is confined to a fixed high-symmetry point away from $\KK$ rather
than to the gap itself: at $\Gamma$ ($k_x=0$), $w_0$ of the same,
continuously tracked upper branch dips sharply below the Dirac-point
value, with the dip deepening monotonically with coupling across the
entire range studied, from $w_0(\Gamma)=0.0035$ at $A_0/c=0.01$ to
$0.0650$ at $A_0/c=0.05$---consistent with an ordinary avoided crossing
whose splitting scales with $A_0$, at a location set by the band
structure rather than by the cavity. 
The overall picture is therefore a Haldane mass generated in the matter sector at $\KK$, a
mechanism consistent with the staggered-flux/photon-valley-locking
picture derived analytically for chiral-cavity graphene by Yang and
Jiang~\cite{Yang2025}, coexisting with, but mechanistically distinct
from, a handful of ordinary photon-replica avoided crossings elsewhere
in the zone whose strength simply tracks $A_0$.

\begin{figure*}[t]
\centering
\includegraphics[width=0.75\textwidth]{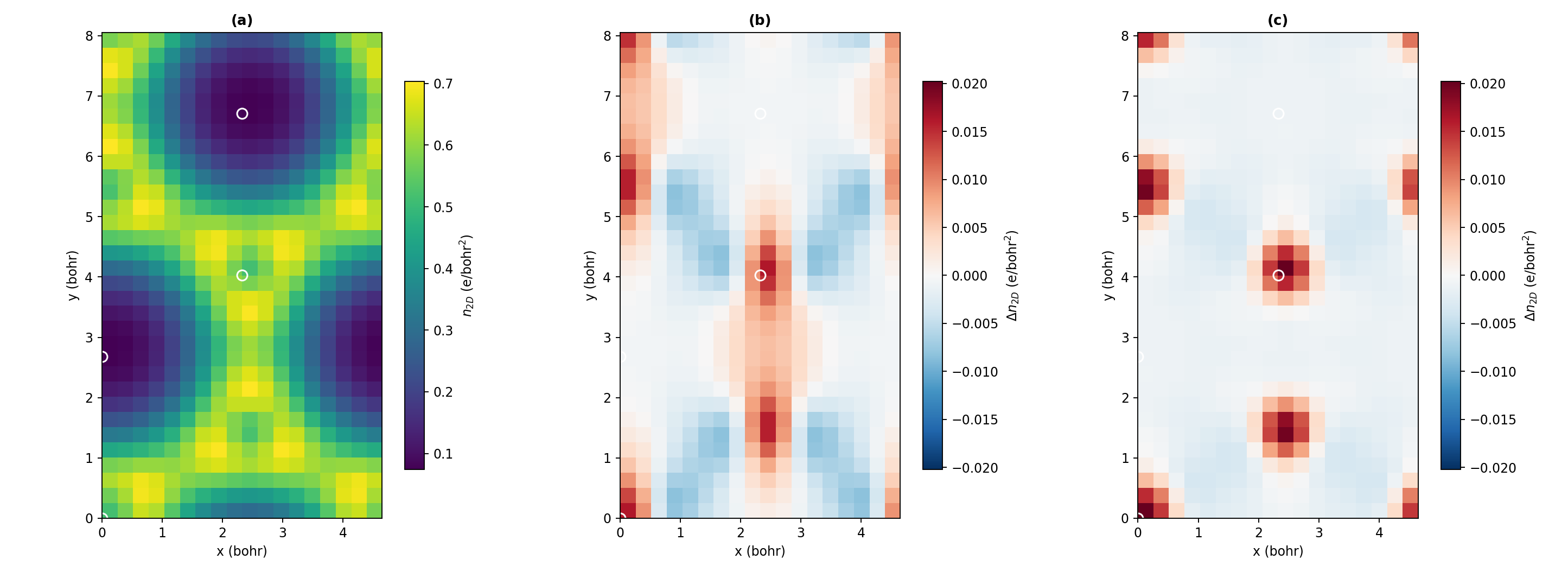}
\caption{Cavity-induced density change at $A_0/c=0.04$, linear color
scale. (a) Cavity-free density $n_{2D}(x,y)$, $z$-integrated. (b)
Linear-cavity minus cavity-free difference $\Delta n_{2D}$ (blue/red =
depletion/accumulation), $\max|\Delta n_{2D}|=1.619\times
10^{-2}\,e/\text{bohr}^2$. (c) Chiral-cavity minus cavity-free
difference, same scale, $\max|\Delta n_{2D}|=2.024\times
10^{-2}\,e/\text{bohr}^2$. White circles mark the four carbon atoms.}
\label{fig:density}
\end{figure*}

\begin{figure*}[t]
\centering
\includegraphics[width=0.75\textwidth]{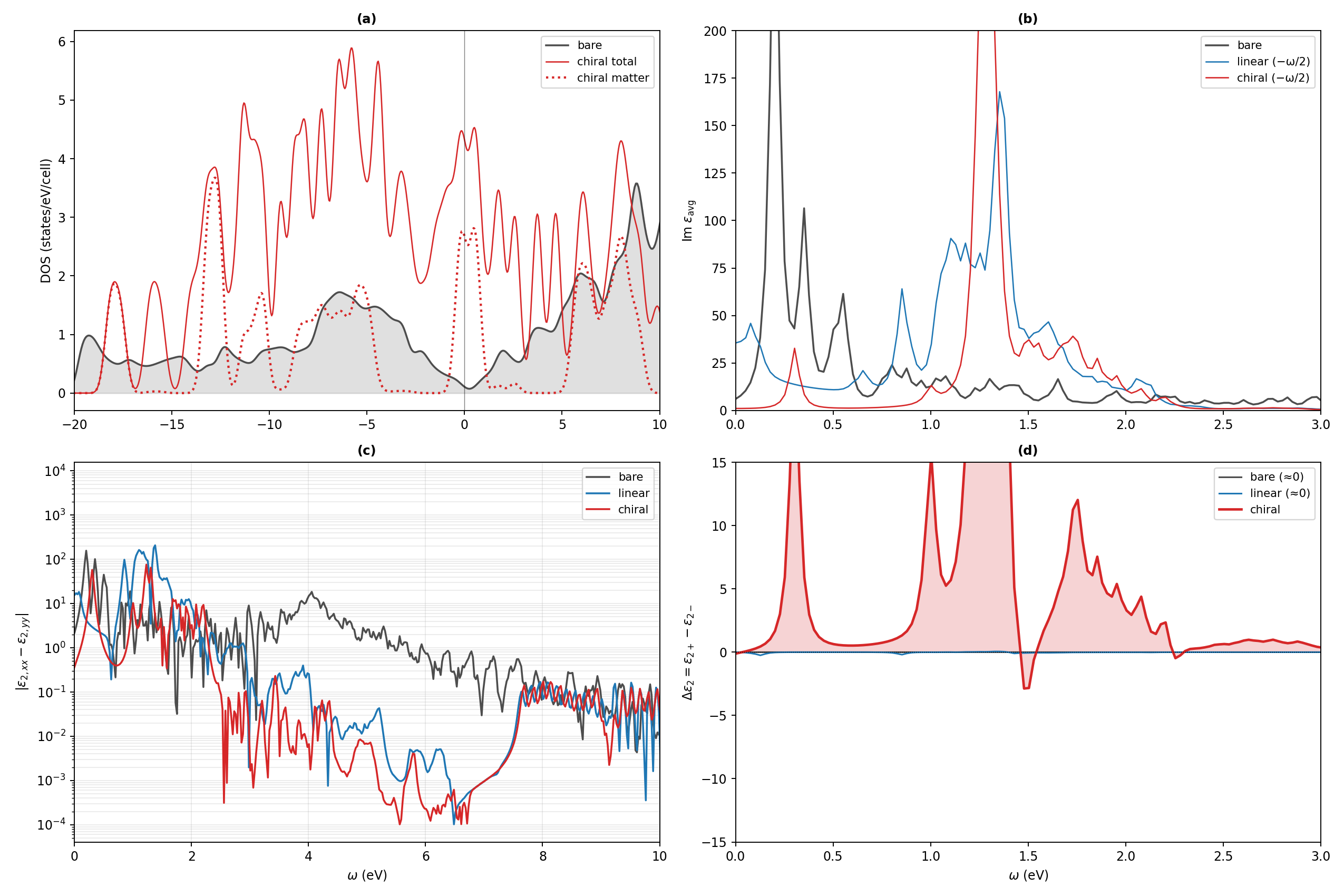}
\caption{Density of states and dielectric response at $A_0/c=0.04$.
(a) DOS; ``total'' sums all four Fock sectors, ``matter'' is the
$w_0$-weighted curve. (b) Absorption $\mathrm{Im}\,\bar\varepsilon(\omega)$.
(c) Linear dichroism $|\varepsilon_{2,xx}-\varepsilon_{2,yy}|$. (d)
Circular dichroism $2\,\mathrm{Im}\,\varepsilon_{2,xy}$.}
\label{fig:dos}
\end{figure*}

We have also calculated the density of states (Fig.~\ref{fig:dos}(a))
and the real-space electron density (Fig.~\ref{fig:density}) in the
chiral, linear, and cavity-free cases. Summed over all four Fock
sectors, the total density of states shows many more peaks than the
bare curve: each matter band spawns its own polariton and
photon-number-replica states displaced by multiples of $\omega$, and
summing over all of them redistributes the smooth bare curve into a
comb of sharp sub-peaks spread across the entire energy window. It
integrates to $55.5$ below $E_F$ for the chiral case at $A_0/c=0.04$,
far more than the $16.0$ expected for $16$ valence electrons; a
$w_0$-weighted, matter-projected density of states, by contrast,
tracks the main spectral features of the bare curve and recovers
$16.0$. The elevated total is not a violation of particle-number
conservation: it counts polariton quasiparticles, most of which carry
substantial photonic character and are not occupied electronic states
in the ordinary sense, whereas the matter-projected curve is the
physically comparable quantity. At $E=0$ the chiral density of states
is genuinely occupied; the bare density of states, by contrast, is
expected to vanish there at the gapless Dirac point and is nonzero in
the plot only because of the finite smearing width used to broaden the
calculation onto a continuous curve. The electron density
itself integrates to $N_e=16.0000$ for the cavity-free, linear, and
chiral densities alike.
Both cavities push charge from the interstitial
region onto the C--C bonds (Fig.~\ref{fig:density}), with a
cavity-induced change $\max|\Delta n_{2D}|\simeq1.6$--$2.0\times
10^{-2}\,e/\text{bohr}^2$ out of a total areal density of order
$0.4$--$0.5\,e/\text{bohr}^2$. The linear and chiral difference
patterns are of comparable magnitude but differ in detail, reflecting
the different symmetry of the two drives: the linearly polarized field
singles out a fixed in-plane axis, while the chiral field respects the
threefold rotational symmetry of the honeycomb lattice on average;
further detail is given in the Supplemental Material~\cite{supp}.

The dielectric tensor (Fig. \ref{fig:dos}) is evaluated in the independent-particle (RPA)
approximation from the polaritonic band structure. The linear and chiral
cavities produce distinct optical fingerprints, shown in
Fig.~\ref{fig:dos}(b)--(d). Panel (b) shows the isotropic absorption
$\mathrm{Im}\,\bar\varepsilon(\omega)$, with the linear and chiral
curves plotted at $\omega-\omega/2$: the same $-\omega/2$ offset that
enters the Jaynes--Cummings gap formula, $\Delta=\sqrt{(\omega/2)^2+
\lambda^2}-\omega/2$, above, removed here so that the polariton
absorption edge is compared to the bare curve without the generic
half-photon offset common to any Jaynes--Cummings doublet; panels (c)
and (d), by contrast, are plotted on the physical, unshifted frequency
axis. Both cavities suppress and shift the bare curve's strong low-frequency
peak---the interband absorption edge of the near-gapless Dirac
cone---to higher energy, by different amounts and with different
lineshapes for the two polarizations, consistent with the
polarization-dependent renormalization reported above. The linear
cavity, whose fixed polarization axis breaks the in-plane isotropy,
generates a strong linear dichroism
$|\varepsilon_{2,xx}-\varepsilon_{2,yy}|$ [Fig.~\ref{fig:dos}(c)] with
essentially vanishing circular dichroism. The chiral cavity does the
opposite---it preserves $\varepsilon_{xx}\simeq\varepsilon_{yy}$ but
generates a circular dichroism $\varepsilon_{2+}-\varepsilon_{2-}$ of
order $5$--$10$ [Fig.~\ref{fig:dos}(d)], the
direct optical signature of the broken time-reversal symmetry of the
cavity vacuum, changing sign through the gap region; the off-diagonal
$\varepsilon_{xy}$, proportional to the anomalous Hall conductivity, is
likewise nonzero only in the chiral case. This circular dichroism is
the measurable, polarization-resolved fingerprint of the Haldane mass
reported above: a transmission or reflection measurement resolved by
circular polarization would see the two cavities' distinct signatures
directly, without requiring a transport probe of the (much smaller,
and experimentally more demanding) anomalous Hall response itself.

The same symmetry distinction leaves an independent, purely kinematic
fingerprint in the polariton dispersion itself, away from the gap: the
local Fermi velocity at the folded Dirac point, resolved along the two
inequivalent in-plane cuts through it, $k_x$ (the $\Gamma$--$X$
direction of Fig.~\ref{fig:bands}) and $k_y$ (transverse to it). The
bare cone is itself mildly anisotropic, $v_\perp/v_\parallel-1\approx
5\%$, an intrinsic trigonal-warping-scale effect of the rectangular
cell rather than a cavity artifact; full fitting details are given in
the Supplemental Material~\cite{supp}. Table~\ref{tab:vF} gives the
cavity-induced anisotropy at every coupling studied. The chiral
cavity renormalizes the cone almost isotropically, tracking within a
few percent of its bare shape across the whole range even as its
overall scale drops by a third; the linear cavity does the opposite,
distorting the cone with a magnitude that grows monotonically with
coupling and crosses sign near $A_0/c\approx0.025$. The same fixed
in-plane axis responsible for the linear dichroism above also imprints
itself kinematically on the cone itself, while the chiral cavity's
circularly symmetric coupling preserves the cone's shape even as it
renormalizes its overall scale---a second, independent confirmation of
the symmetry-complementary picture.

\begin{table}[t]
\centering
\caption{Cavity-induced Fermi-velocity anisotropy
$(v_\perp-v_\parallel)/v_\parallel$ at the folded Dirac point, relative
to a bare value of $+4.9\%$.}
\begin{tabular}{ccc}
\hline\hline
$A_0/c$ & linear & chiral \\
\hline
0.01 & $+3.8\%$ & $+5.2\%$ \\
0.02 & $+1.9\%$ & $+6.0\%$ \\
0.03 & $-1.1\%$ & $+7.4\%$ \\
0.04 & $-4.8\%$ & $+7.3\%$ \\
0.05 & $-9.8\%$ & $+7.4\%$ \\
\hline\hline
\end{tabular}
\label{tab:vF}
\end{table}

\textit{Topological characterization.}---The Haldane picture above is
inferred from the local gap character at $\KK$; we now confirm it
directly with a genuine Brillouin-zone topological invariant. We
compute the Chern number of the occupied manifold---the eight
polaritonic bands adiabatically connected to the eight cavity-free
valence bands of the four-atom cell, a fixed-dimension manifold rather
than one reselected at each coupling by an energy window or
photon-vacuum-weight threshold (Supplemental Material~\cite{supp} states
this definition precisely)---using the lattice
(Fukui--Hatsugai--Suzuki) construction~\cite{Fukui2005}: the $U(1)$
link variable between neighboring $\kk$-points is built from the
determinant of the occupied-band overlap matrix, and the flux through
each plaquette of the mesh is summed to give $C$, a quantity that is
by construction invariant under any unitary mixing of the occupied
subspace and therefore free of the gauge ambiguity that a
single-band or single-doublet Berry-phase calculation would carry.
Unlike the open $\kk$-path window of Fig.~\ref{fig:bands}, this
requires a mesh that closes exactly on the Brillouin-zone torus. 
Full details, including a validation of the
implementation against an analytically solvable lattice model, are
given in the Supplemental Material~\cite{supp}.

\begin{figure}[t]
\centering
\includegraphics[width=\columnwidth]{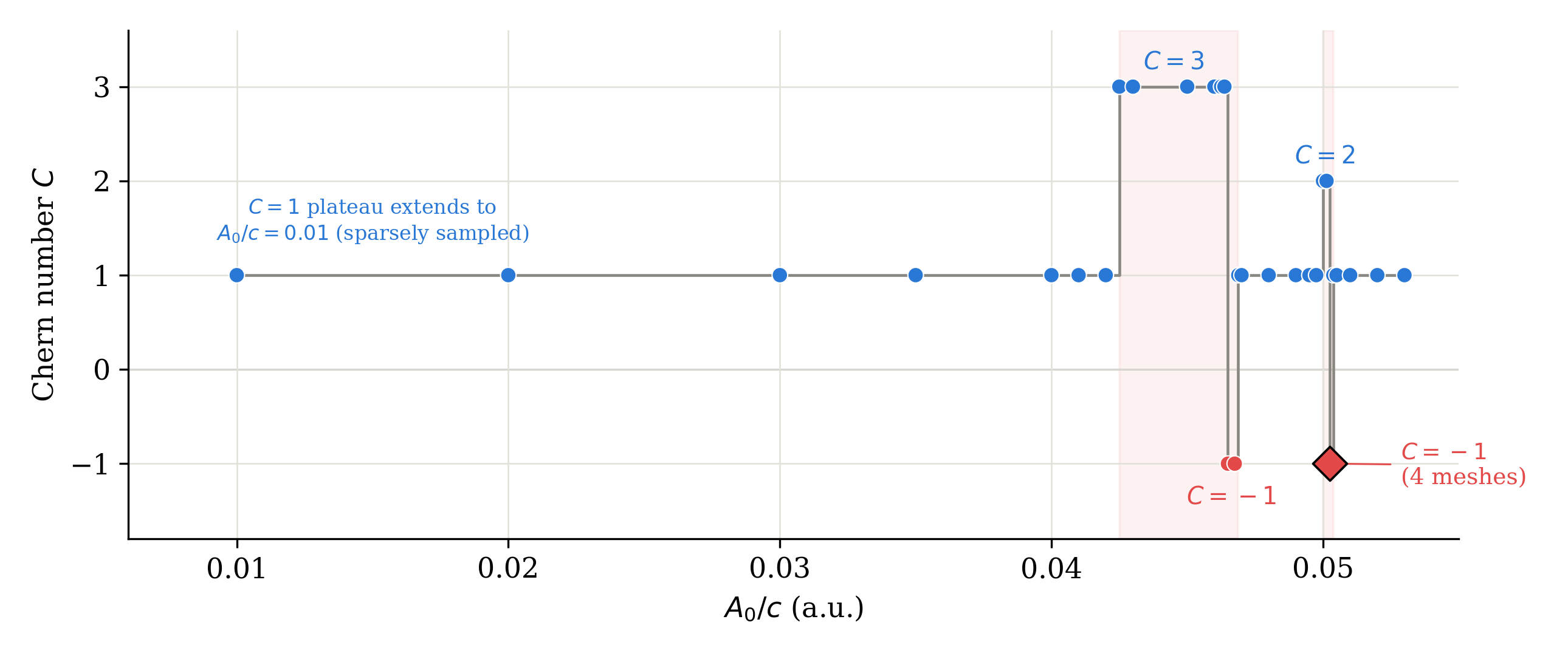}
\caption{Chern number of the occupied eight-band manifold, chiral
cavity, across 29 independently converged couplings. Blue markers:
$C=1,2,3$; red markers: the two sign-reversed $C=-1$ notches, each
$\lesssim1\times10^{-4}$ wide in $A_0/c$ and each immediately preceded
by a jump to a higher-$C$ plateau (shaded). The red diamond at
$A_0/c=0.05025$ is independently confirmed on four meshes spanning
60--9880 $\kk$-points (Supplemental Material~\cite{supp}). Berry-curvature
maps for representative couplings are shown in Fig.~\ref{fig:chernmaps}.}
\label{fig:chern}
\end{figure}

As a first check, the cavity-free case gives a total flux
$\Sigma F/2\pi=0$ to within $3\times10^{-5}$ at every coupling---not
itself a genuine topological invariant, since the gapless cavity-free
manifold does not admit one in the usual insulating sense, but a
useful null test: the two Dirac points, folded onto $k_x=\frac13$ and
$\frac23$ in this cell, each contribute a curvature of $\pm0.4997$
concentrated in a single plaquette---the expected canceling
half-vortices of a time-reversal-symmetric gapless semimetal, and a
nontrivial confirmation that the periodic wraparound links are working
correctly before trusting the gapped result. For
the chiral cavity the manifold is gapped everywhere on the mesh we
studied, and $C$ is quantized cleanly at every one of 29 couplings
surveyed between $A_0/c=0.01$ and $0.053$---but far from a simple,
monotonically increasing sequence. Figure~\ref{fig:chern} shows the
result: $C=1$ from $A_0/c=0.01$ up to $0.042$, then a transition to
$C=3$ persisting to $0.0464$, then, within a window only
$\sim1\times10^{-4}$ wide in $A_0/c$, a sign-reversed notch to $C=-1$
before recovering to $C=1$. The pattern repeats at higher coupling: a
second transition to $C=2$ at $A_0/c=0.05$ is followed, again within
$\sim1\times10^{-4}$, by a second $C=-1$ notch before the manifold
settles back to $C=1$ up to the largest coupling studied. Both
notches follow the same motif---jump to a higher plateau, sign-reversed
dip, recovery to $C=1$---and the deeper of the two, at
$A_0/c=0.05025$, is confirmed on four independently generated meshes
of 60, 1450, 5800, and 9880 $\kk$-points 
(Supplemental Material~\cite{supp}). We
additionally cross-checked this staircase by an independent
Wilson-loop (hybrid Wannier-center) calculation~\cite{Yu2011}
which reproduces $C$ at every coupling studied 
(Supplemental Material~\cite{supp}).

\begin{figure*}[t]
\centering
\includegraphics[width=0.7\textwidth]{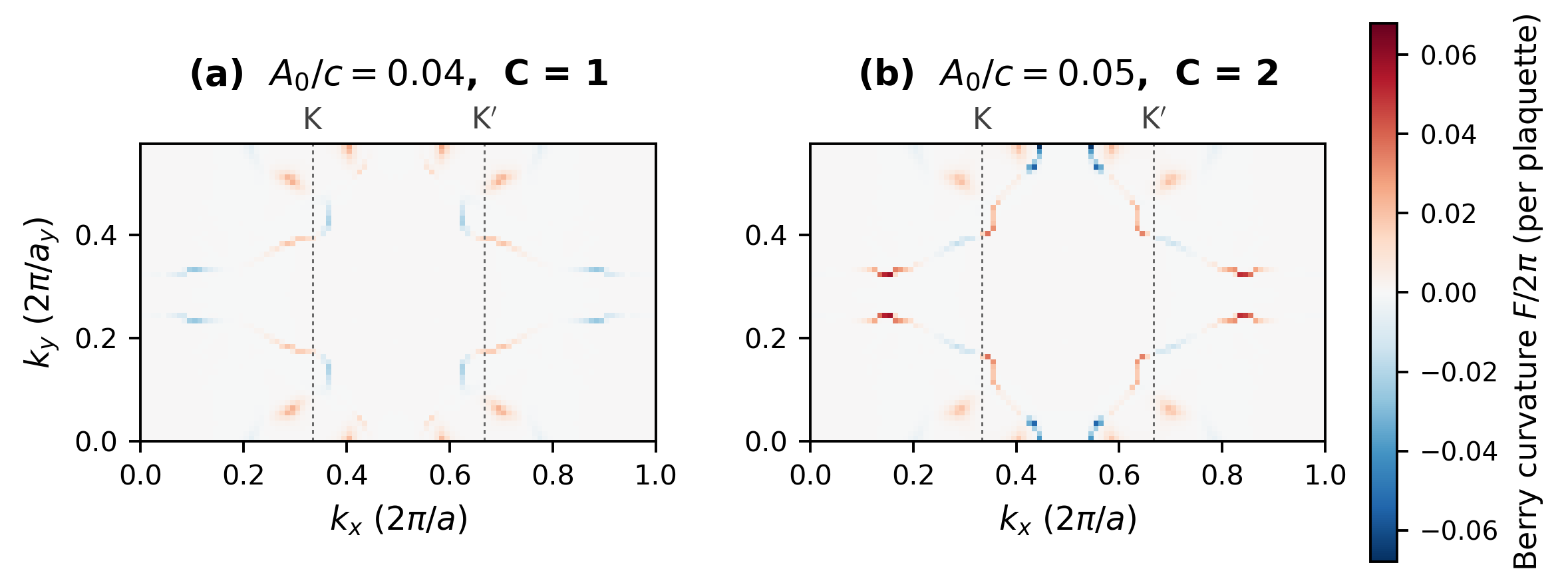}
\caption{Berry curvature of the occupied eight-band manifold over the
full Brillouin zone, chiral cavity, at (a) $A_0/c=0.04$ ($C=1$) and (b)
$A_0/c=0.05$ ($C=2$), the two original anchor couplings. Dotted lines
mark the folded Dirac points $\KK,\KK'$ ($k_x=\frac13,\frac23$).
Further detail is given in the Supplemental Material~\cite{supp}.}
\label{fig:chernmaps}
\end{figure*}

Figure~\ref{fig:chernmaps} maps this curvature over the full zone at
the two original anchor couplings. At both, the curvature avoids the
folded Dirac points $\KK,\KK'$ themselves, consistent with the
matter-like character of the gap-forming states found there in
Fig.~\ref{fig:bands}; instead it concentrates at the same off-symmetry
avoided crossings identified in the band-structure discussion above.
Between the two couplings a new pair of hotspots appears directly on
the $\KK$--$\KK'$ line (near $k_x\approx0.44,0.56$, $k_y\approx0.57$),
the real-space signature of the additional avoided crossing
responsible for the $C=1\to2$ transition at $A_0/c=0.05$: the
topological transition is thus traceable to a concrete, localized
feature of the curvature map rather than to a global redistribution
over the zone.

That $C$ changes at all between couplings otherwise connected by a
smooth, monotonic Haldane-gap-scaling law at $\KK$
(Table~\ref{tab:scaling}) is not a contradiction: $C$ is a discrete
topological invariant that can only change when the occupied manifold
becomes gapless somewhere in the zone, so each transition in
Fig.~\ref{fig:chern} demands its own bulk gap-closing and reopening
event away from $\KK$ itself, where the Haldane gap never closes over
this range. What the full survey adds to a simpler two-point picture
is that these events are neither smooth nor uniformly signed: the
top-20-plaquette share of the total curvature---a diagnostic of how
localized in $\kk$-space a given transition is---rises from a clean
$13$--$15\%$ baseline away from any transition to $25$--$37\%$ inside
both narrow windows (Supplemental Material~\cite{supp}), confirming
each notch is a genuinely sharp, localized band-touching event rather
than a numerical artifact; yet the jumps involved ($+2,-4,+2,+1,-3,+2$
in sequence) are not the simple $+1$-per-crossing ladder that
independently turning-on, chirality-aligned photon-replica crossings
would predict. The overall picture is therefore a Haldane mass
generated and topologically anchored at $\KK$, punctuated by a sparse
set of much narrower, sign-varying gap-closing events elsewhere in the
zone as the coupling is tuned---evidence that a chiral cavity can
drive a two-dimensional Dirac material through a genuinely
non-monotonic sequence of Chern phases, including sign-reversed
windows, rather than a simple ladder of increasing $C$. This
quantized, sign-changing Chern number is the topological invariant
underlying the circular dichroism reported above: a nonzero $C$ is
possible only with the broken time-reversal symmetry that also
produces the chiral-cavity-only $\varepsilon_{2+}-\varepsilon_{2-}$
dichroism of Fig.~\ref{fig:dos}(d), tying the band-structure-level
Haldane picture and the full-zone topological invariant to the same
measurable optical signature.

\textit{Conclusion.}---We have presented a plane-wave pseudopotential
formulation of cavity QED-DFT that requires no new machinery beyond a
crystal-momentum shift by the quantized vector potential, and used it
to show that a chiral cavity drives monolayer graphene through a
polarization-selective Haldane gap, confirmed by a quantized,
non-monotonic Chern-number staircase ---while a linear cavity leaves
the Dirac point gapless; full formalism, a self-consistency
instability we identified and resolved, and further details are given
in the Supplemental Material~\cite{supp}. Because
the method reuses existing
plane-wave infrastructure almost unchanged, it extends naturally to
any system already accessible to plane-wave DFT: multilayer and
moir\'e structures, where cavity engineering of twist-angle physics has
been proposed and, in the terahertz regime, experimentally
approached~\cite{M.Kulkarni2026,Tay2025,Yang2025,1tlw-g26r,8qx2-xxh2};
topological and spin--orbit-coupled materials, where cavity-induced
topological phase transitions have very recently been demonstrated
first-principles in bulk HgTe~\cite{Shin2026sciadv}; and, with a
multimode generalization, cavities supporting several photon modes or a
photonic continuum.

\section{Acknowledgment}
This work was supported by the National Science Foundation (NSF) under
Grant No. DMR-2217759.

\section*{Data Availability Statement}
All data and code are available at https://github.com/kvvandy/tddft.


%

\end{document}